\documentclass[a4paper,11pt]{article}
\pdfoutput=1 % if your are submitting a pdflatex (i.e. if you have
\usepackage{jinstpub} % for details on the use of the package, please
\usepackage{subfigure}

\title{\boldmath Determination of impact parameter for CEE with digi-input neural networks}

\author[a]{Botan Wang,}
\author[a,1]{Yi Wang,\note{Corresponding author.}}
\author[a]{Dong Han,}
\author[b]{Zhigang Xiao}
\author[c]{and Yapeng Zhang}

\affiliation[a]{Key Laboratory of Particle and Radiation Imaging, Department of Engineering Physics,\\Tsinghua University, Beijing 100084, China}
\affiliation[b]{Department of Physics,\\Tsinghua University, Beijing 100084, China}
\affiliation[c]{Institute of Modern Physics,\\Chinese Academy of Science, Lanzhou 730000, China}

\emailAdd{yiwang@mail.tsinghua.edu.cn}

\abstract{The impact parameter characterizes the centrality in nucleus-nucleus collision geometry. The determination of impact parameters in real experiments is usually based on the reconstructed particle attributes or the derived event-level observables. For the scheduled Cooler-storage-ring External-target Experiment (CEE), the low beam energy reduces correlation between the impact parameter and charged particle multiplicity, which decreases the validity of the explicit determination methods. This work investigates a few neural network-based models that directly take the digitized signals from the external Time-of-flight detectors as input. The model with the best performance shows a mean absolute error of 0.479 fm with simulated U-U collisions at 0.5 AGeV. The performances of the models implemented with digi inputs are compared with reference models with phase space inputs, showing the capability of neural networks to handle the original but potentially interrelated digitized signal information.}


\arxivnumber{2307.15355} % only if you have one

\begin{document}
\maketitle
\flushbottom

\section{Introduction}
\begin{figure}[htbp]
\centering % \begin{center}/\end{center} takes some additional vertical space
\includegraphics[width=.5\textwidth]{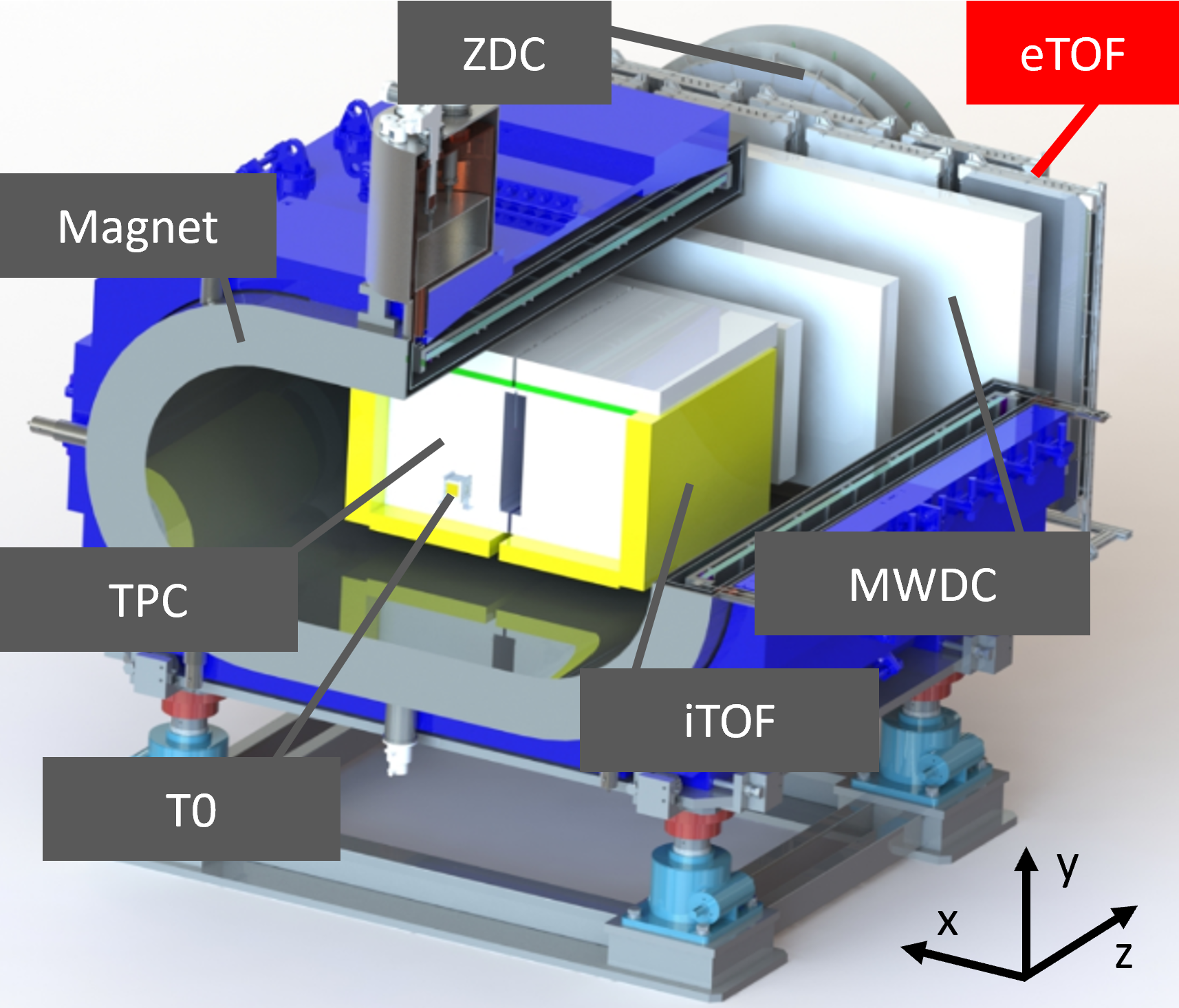}
\caption{\label{fig:1} Detector layout of CEE.}
\end{figure}

\begin{figure}[htbp]
\centering
\includegraphics[width=.7\textwidth]{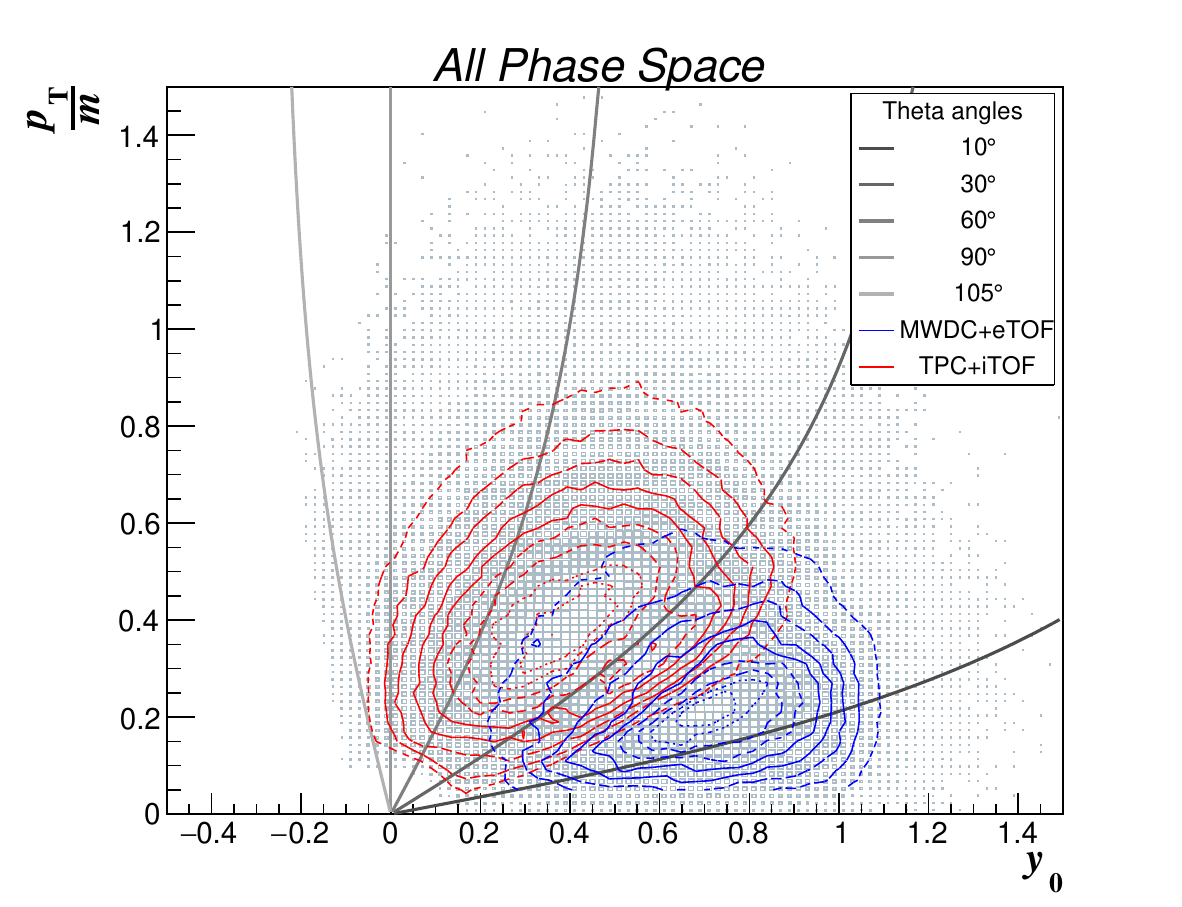}
\caption{\label{fig:2} Phase space distribution of CEE. The size of the pixels stands for the particle density, and the contours show the coverage of the forward and transverse detectors.}
\end{figure}

Relativistic nucleus-nucleus collision is one of the main approaches to studying the Equation of State (EoS) of the nuclear matter\cite{jacob_quark_1982}. By designing the collision system and energy, the compressed matter is expected to reach a specific state in the Quantum Chromodynamics (QCD) phase diagram\cite{gupta_scale_2011}. Although such a state evolution lasts only for tens of fm/$c$, results and evidence on physics issues such as critical point of phase transition\cite{luo_search_2017}, Quark-Gluon Plasma\cite{jacak_exploration_2012}, color-superconductivity\cite{alford_color_2008} can be made through the detection and analysis of the final-state particles. There are many colliders\cite{ackermann_star_2003,noauthor_alice_2008,kim_detector_2019} or fixed-target facilities\cite{meehan_fixed_2016,galatyuk_hades_2014} around the world whose detectors are built for the acceptance of collision products such as charged particles, neutrons, gamma rays, and projectile fragments. The Cooler-storage-ring External-target Experiment (CEE)\cite{lu_conceptual_2017} is a fixed-target heavy-ion experiment running in an energy range of 0.5-1.2 AGeV. The time-of-flight (TOF) technique\cite{wang_multigap_2020}, which requires the cooperation of tracking and timing detectors, is practiced for charged particle identification. Figure~\ref{fig:1} shows the detector layout of the CEE. Since there is a considerable variation of particle momentum with polar angle, two arrays of detectors are designed for identifying the particles at the front ($\theta$<25°) and at large angles ($\theta$>25°). The Time Projection Chamber (TPC)\cite{huang_laser_2018} and the inner TOF (iTOF)\cite{wang_cee_2022} systems are installed inside the dipole magnet and cover the intermediate rapidity zone. As front-angle detectors, the Multi-Wire Drift Chamber (MWDC)\cite{yi_prototype_2014} and the external TOF (eTOF)\cite{wang_cee-etof_2020,wang_external_2023} wall are placed downstream of the magnet, allowing longer flight distance for better resolution in particle identification. Figure~\ref{fig:2} shows the phase space distribution, i.e., the particle distribution as a function of the reduced rapidity $y_0$ and the reduced transverse momentum $p_{\rm T}/m$, where the acceptances of the two detector arrays can be recognized. A T0 detector\cite{hu_extensive_2020} is also included for start time measurements. A Zero-Degree Calorimeter (ZDC)\cite{liu_event_2023} is placed behind the eTOF to measure the baryons used for the reconstruction of event plane and centrality.

Given a fixed beam energy, centrality is the most descriptive geometric quantity determining a collision's initial state. An equivalent description of centrality is the impact parameter $b$, defined as the perpendicular distance between the trajectories of both nuclei before their collision. Central collisions result in a large number of interacting nucleons, known as participants, and compressed nuclear matter with high density. There are several EoS-related research topics, such as the collective and elliptical flow and symmetry energy, which need the recognition of the central, semi-peripheral, and peripheral collisions\cite{goyal_impact_2013,kaur_elliptical_2011}. The most common method in experiments is to measure the distribution of a single observable with a strong correlation with the impact parameter. Then, the distribution can be processed by a sharp cut-off or Bayesian methods, which give discrimination strategies\cite{cavata_determination_1990,miller_glauber_2007,frankland_model_2021,das_relating_2018}. Such observables include the total charged multiplicity, the transverse momentum, the number of intermediate-mass fragments, the total number of participant neutrons, etc. The fruitful collection of the above features also inspired the multivariate analyses, which turned the centrality determination into a typical instance of feature recognition. Since the 1990s, there have been works that apply Support Vector Machine (SVM)\cite{de_sanctis_classification_2009}, Gradient Boosting Decision Tree (GBDT)\cite{li_application_2020}, and several kinds of neural networks\cite{haddad_impact_1997,david_impact_1995,omana_kuttan_fast_2020} to the regression of the impact parameter. Such works have two main characteristics: the regression or classification model is trained with simulated data, which can provide the target ‘TRUE’ impact parameter, and the input variables to the models are the highly correlated observables obtained through event reconstruction. 

The development of Artificial Intelligence and Deep Neural Networks (DNN) has expanded the capabilities of regression models in non-linear tasks and made it possible for models to handle data with primitive features. In recent works on centrality determination, tracks and hits of the final state particles extracted from the detectors have been considered direct features. \emph{F. Li et al.} translated the particle production into a phase space image in a 30$\times$30 shape of pixels and implemented a Convolution Neural Network (CNN)\cite{krizhevsky_imagenet_2017} based model to regress the impact parameter\cite{li_application_2021}. Then, they tested the precision under the Au-Au collision data in an intermediate low beam energy range of 0.2-1.0 AGeV. \emph{M. O. Kuttan et al.} were dedicated to developing a CNN-based model called PointNet, which could handle the variable length and commutable input of hits and tracks\cite{omana_kuttan_fast_2020}. The whole structure was applied to the regression of the impact parameter in Au-Au collisions at 10 AGeV. 

The above works inspired the impact parameter determination of CEE. We propose the regression with DNNs using the channel response of the eTOF detectors as input. Tests showed that using data from the detector leads to more accurate predictions of the impact parameter compared to using common event-level features like phase space distribution. This paper is organized as follows: Section~\ref{sec:2} describes the task of impact parameter determination in the context of CEE, provides basic information about the eTOF wall, and accounts for how the features from the eTOF detectors are extracted. Section~\ref{sec:3} describes how the simulation is designed and carried out, including the data preparation and the DNN models. Section~\ref{sec:4} is dedicated to the discussion of results. A summary is made in the last section.

\section{Description of the task}
\label{sec:2}
The typical collision system for CEE is U-U at 0.5 AGeV. The collision energy is below the required threshold for most mesons and resonances, leaving most charged products as baryons like protons, deuterons, tritons, and heavier ions\cite{wang_external_2023}. As a result, event-by-event fluctuations for central collisions are high, which limit the centrality dependence on multiplicity\cite{omana_kuttan_fast_2020}. Given the expectation of high prediction uncertainty in the CEE condition, many strategies for centrality determination have been considered, such as using the angle distribution of energy deposition in ZDC or reconstructed tracks of the entire system. It is expected that strategies based on different detectors as data sources can be combined into a decisive final predictor.

\begin{figure}[htbp]
\centering
\includegraphics[width=.5\textwidth]{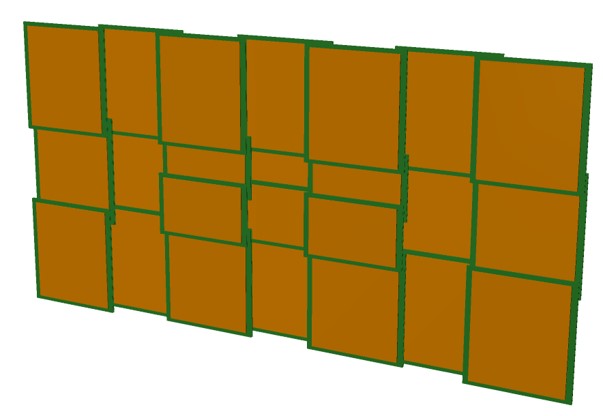}
\caption{\label{fig:3} Detector layout of the eTOF wall.}
\end{figure}

The eTOF wall is one of the best detector candidates for centrality determination thanks to the extensive coverage (3.2$\times$1.6 m$^2$), high granularity (1344 channels in total), and fast signal shaping. The eTOF wall comprises 24 Multigap Resistive Plate Chambers (MRPC) with 10 gas gaps of 0.25 mm thickness. Figure~\ref{fig:3} shows the detector layout of the eTOF wall, where the MRPCs cover an active area of 3.2$\times$1.6 m$^2$. The eTOF wall contains 672 horizontal strips, read out from both ends. They are in the same size of 48$\times$1.5 cm$^2$ and arranged in an interval of 1.7 cm in every detector. The 6 smaller MRPCs in the inner region of the eTOF wall have 16 readout strips, while the others have 32. The eTOF wall is divided into 7 columns, which results in a \textbf{96$\times$7 layout} of readout strips. The primitive signals, induced onto the readout strips by the avalanches in MRPC’s gas gaps, are amplified and discriminated by the NINO-based\cite{anghinolfi_nino_2004} Front-End Electronics (FEE)\cite{lu_readout_2021}. The output signals are recorded by the Time-to-Digital Module\cite{lu_readout_2021} for their leading and trailing time when the amplified signals cross a threshold (i.e., 150 mV for eTOF FEEs). The expected efficiency and time resolution of the eTOF MRPCs are better than 95\% and 60 ps, respectively, and have been verified on real-size prototypes in the cosmic and beam tests.

\begin{figure}[htbp]
\centering
\includegraphics[width=.7\textwidth]{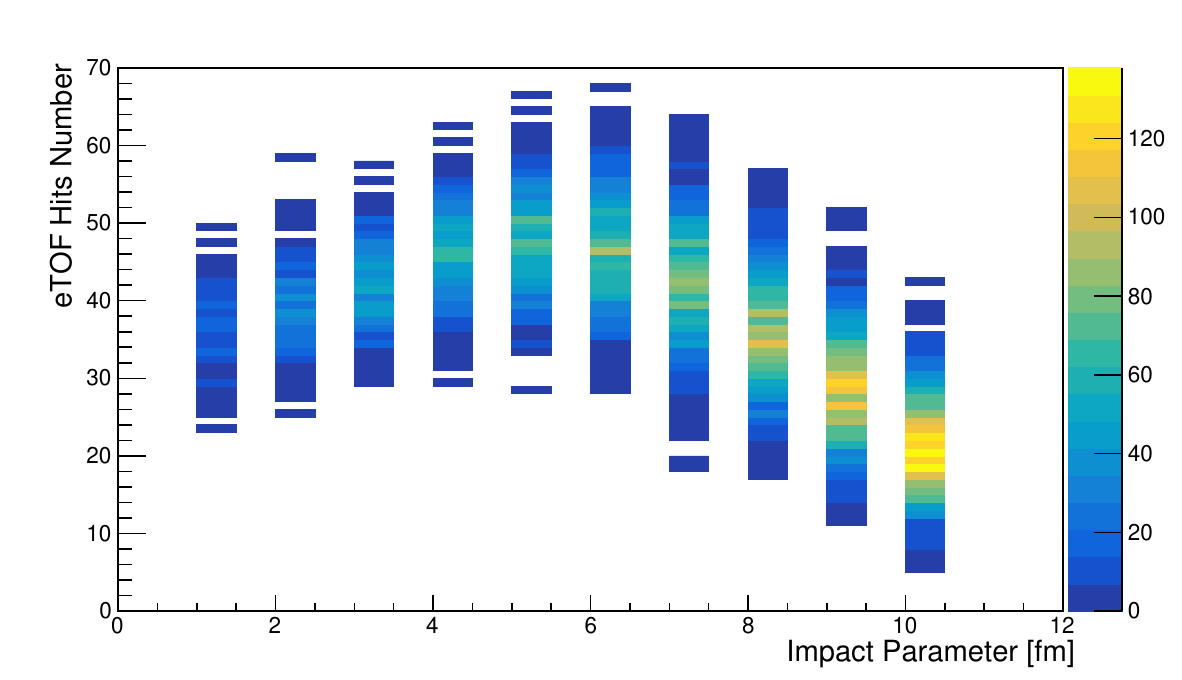}
\caption{\label{fig:4} Multiplicity distribution of eTOF as a function of impact parameter.}
\end{figure}

Due to the low collision energy and limited coverage at frontal angles ($\theta$<25°), there is a complete loss of monotonic relationship between accepted particle multiplicity and impact parameter for the eTOF wall, as illustrated in Figure~\ref{fig:4}. The observed positive correlation between multiplicity and impact parameter, specifically for impact parameters less than 5 fm, is probably due to the particle acceptance influenced by transverse momentum in central collisions. Conversely, the negative correlation for $b$ exceeding 5 fm is resulted from the diminishing number of participants. Therefore, it is essential to consider multiple features when using eTOF as a data source. In other words, not only the accepted hit multiplicity but also their temporal and spatial features should be included. Such features can be easily extracted from the digitized signals, i.e., the recorded leading and trailing-edge times mentioned above. The average time of both ends of the strip represents the hit time, and the differential time represents the hit position along the strip. The channel layout of the eTOF wall represents a gridding of the phase space. Therefore, it is reasonable to believe that the digi-input strategy will not lose the information or the granularity in the context of centrality determination.

This work presents a procedure which takes digitized signals from eTOF and gives event-by-event predictions of impact parameters. However, it is important to maintain the validity of the methods and models when using the experimental signals in future practices. At least two factors should be considered on the quality of the experimental data. The first is the correct event assembly, which requires that the acquired signals are sourced from a single collision. In CEE condition, over 98\% of the in-event charged particles reach the TOF detectors within a time window of 25 ns. Accordingly, the time windows for triggering and data acquisition are 75 ns and 1 us, respectively. Given the interaction rate of 10 kHz, the channel response of eTOF in the same event can be organized correctly. The second factor is the stability of the data distribution, such as the reconstructed time and position. Should the experimental data distribution be biased or unstatic, model prediction errors would result. During the operation of CEE, parameters from both the detectors (signal propagation speed, time resolution, etc.) and the electronics (processing time of readout, triggering, etc.) are stable with time. Given a feature distribution independent from instrument conditions, proper normalization can be carried out to minimize the differences between the simulation and experimental data distributions. The simulation study utilizes ground truth information to evaluate model performances. In practical scenarios where true impact parameters are unavailable, evaluation techniques may leverage alternative physics observations or predicted possibility distribution\cite{haddad_impact_1997,das_relating_2018}. This work leaves such experimental issues for further investigation in the future.

\section{Description of the Simulation}
\label{sec:3}
\subsection{Dataset generation}
Datasets are generated following the procedures below: Firstly, collision data is generated, which gives the produced steady particles and their initial kinetic attributes. The physics model of the collision is Isospin-dependent Quantum Molecular Dynamics (IQMD)\cite{hartnack_modelling_1998}, and the coalescence of nucleons is processed after the IQMD calculation since ion yield is significant in the CEE energy. Then, CeeRoot\cite{noauthor_ceeroot_nodate}, a platform based on the FairRoot\cite{al-turany_fairroot_2012} framework, carries out the simulation, which organizes the collision data together with detector geometry and the Geant4 transport model. Finally, we obtain the dataset for model training and testing with proper pre-processing. 

During the CeeRoot simulation, the digitization method\cite{wang_simulation_2022}, which translates the Monte-Carlo points of eTOF into the digitized signals, is carefully designed according to the experimental conditions. Processes of detector response, such as noise hits, hit cluster size, time resolution, and time delays, have been considered in digitization. Noise hits are added randomly following the typical MRPC noise level of 1 Hz/cm$^2$. The time of noise hits is obtained by sampling the uniform distribution U(0, 100) ns, where 100 ns is the typical time window for event assembly. The noise hit position is sampled from U(-1.5, 1.5) ns, corresponding to the strip length range. Due to the transverse diffusion of the avalanche charge, hits with multiple fired strips should be considered in the simulation. In digitization, such an effect is simulated by sampling a cluster size distribution measured in detector tests. For each digitized signal, the signal leading time $T_{\rm leading}$ is calculated as follows:
\begin{equation}
T_{\rm leading} = t_{\rm hit}+\frac{\lvert x_{\rm end}-x_{\rm hit} \rvert}{v_{\rm prop}}+\Delta t_{\rm ele}+\Delta t_{\rm reso}
\label{eq:0}
\end{equation}
where $t_{\rm hit}$, $x_{\rm hit}$ are the hit time and position, $x_{\rm end}$ is the position of the strip end, $v_{\rm prop}$ is the signal propagation speed along the strip. $\Delta t_{\rm ele}$ is the time delay of the electronics, which is sampled from U(0, 5) ns and constant for each specific channel.  $\Delta t_{\rm reso}$ is sampled from a Gaussian distribution with zero mean and 100 ps standard deviation, representing the time resolution. It is inferred from the data normalization that the time delay effect can be eliminated in the model regression and prediction. This is very helpful because of the complexity of the measurement of channel time delay. As the time-slewing effect has not been implemented in the digitization method, the 100 ps smearing is chosen according to the measurement before the time-slewing correction in the cosmic tests\cite{wang_simulation_2022}.

We simulated 1.21 million events of U-U collisions at 0.5 AGeV. The events are distributed uniformly from impact parameter b=0.0 fm to b=12.0 fm with 0.1 fm step. They are randomly split into the train and test set with a 4:1 ratio. Two datasets are extracted from the test set: the so-called $b{\rm d}b$ test set with 120k events and the uniform test set with all the 242k events in the test set. The $b{\rm d}b$ test set is generated following the collision geometry where the differential cross section ${\rm d} \sigma \propto b {\rm d} b$\footnote{Extracted from the test set, the $b{\rm d}b$ test set contains 17 events with $b=0.1$ fm, 33 events with $b=0.2$ fm, ..., 2k events with $b=12.0$ fm}. It is used to evaluate the overall performances of the models because it is close to the realistic collisions; the uniform test set is used to assess performance as a function of the impact parameter. The train set has 968k events in which 80\% are used for model fitting, and 20\% are used for early-stop decision (see Section~\ref{sec:3-3}) after each training epoch.  

\begin{figure}[htbp]
\centering
\includegraphics[width=.7\textwidth]{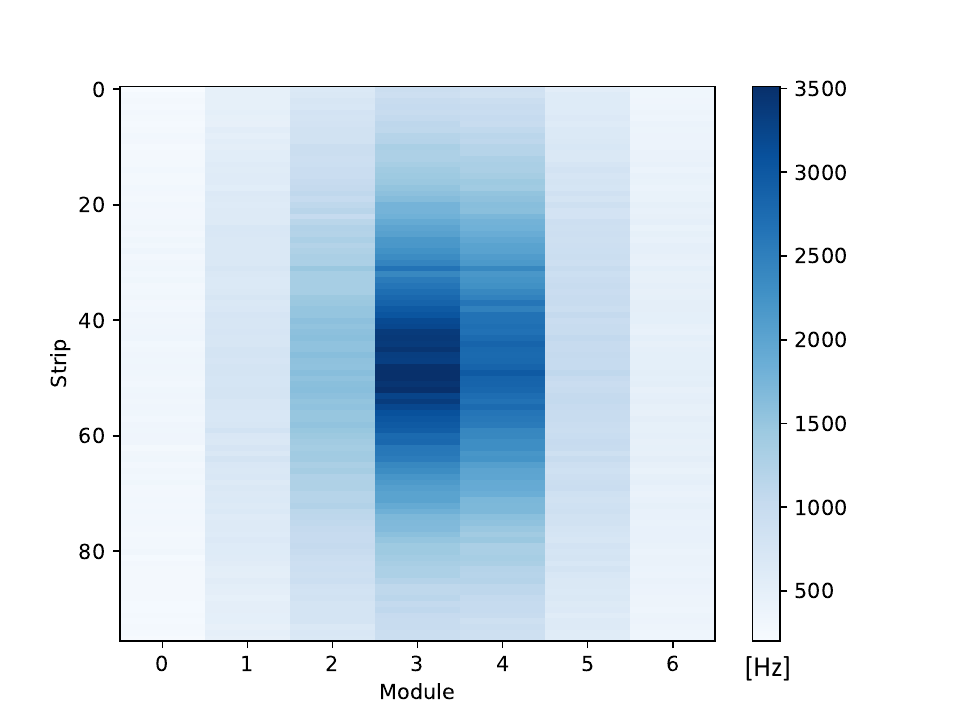}
\caption{\label{fig:5} Counting rate distribution of 672 readout strips for eTOF wall.}
\end{figure}

We collect the features from the total 672 readout strips for each simulated event, including the \textbf{hit flag, hit time, and hit position}. The hit flag is set to 1 if signals are generated from both strip ends. The hit time is calculated as the average of the times recorded at each end, and the hit position corresponds to their half difference. Despite its simplicity, such calculation provides features with explicit physics meaning. The hit flag is set to zero for strips with no hit detected, and the time and position are set to the local average value from the training dataset. The choice of data imputation here is made to obtain zero values after the normalization of the time and position features. The purpose of keeping zero values for the non-fired nodes is to stop the forward propagation of their features implicitly. The first dense layer is designed to exclude the intercept term so that the non-fired nodes with zero input values always result in zero outputs. The absence of a trainable intercept does not affect the capability of the dense layer because the weight on the hit flag plays an equivalent role. Moreover, the zero-value nodes have no function in the attention layer when calculating the linear combination of the nodes with their attributes. Figure~\ref{fig:5} shows the counting rate distribution of the 672 readout strips for the eTOF wall. A descending counting rate is shown due to the collision in fix-target mode.

\begin{figure}[htbp]
  \centering
  \subfigure[]{
      \includegraphics[height=0.25\textwidth]{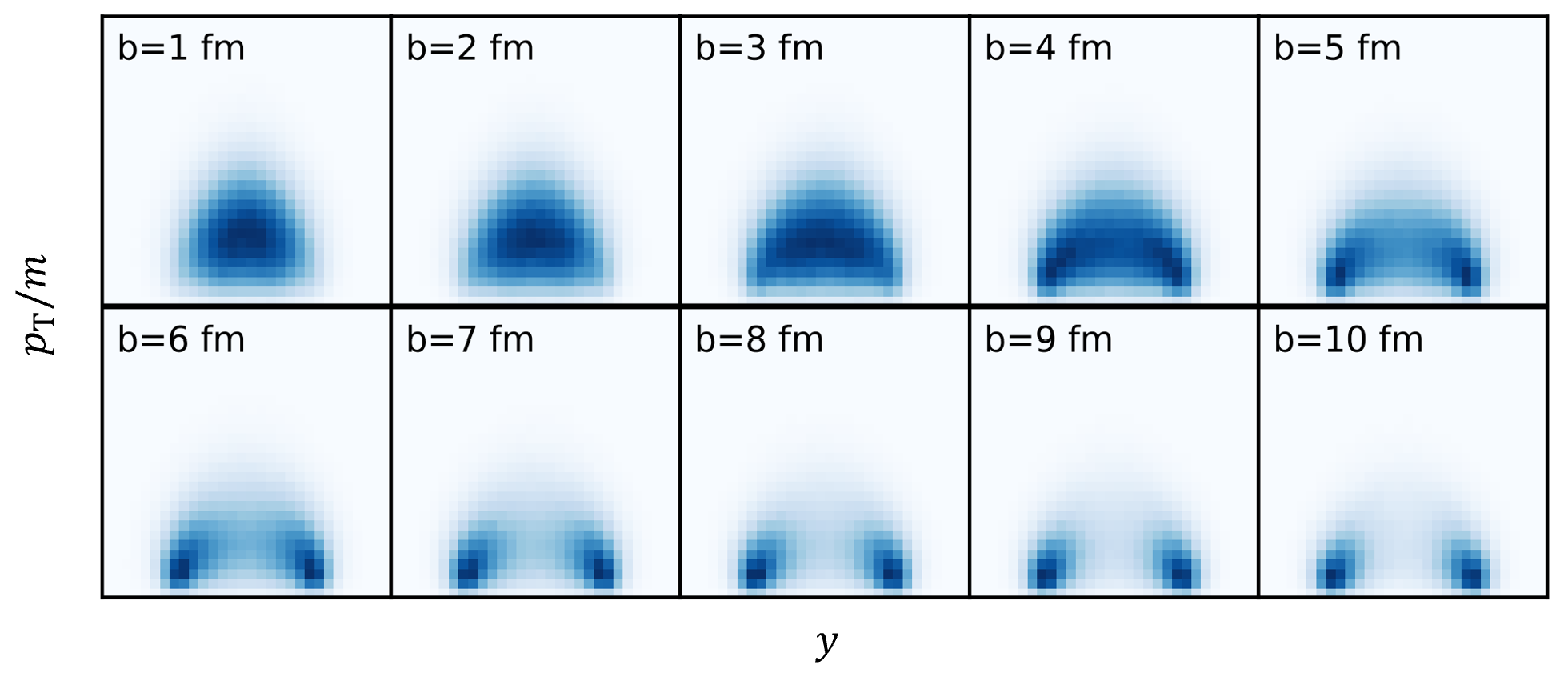}
      \label{fig:6-a}
  }
  % \hfill
  \subfigure[]{
      \includegraphics[height=0.25\textwidth]{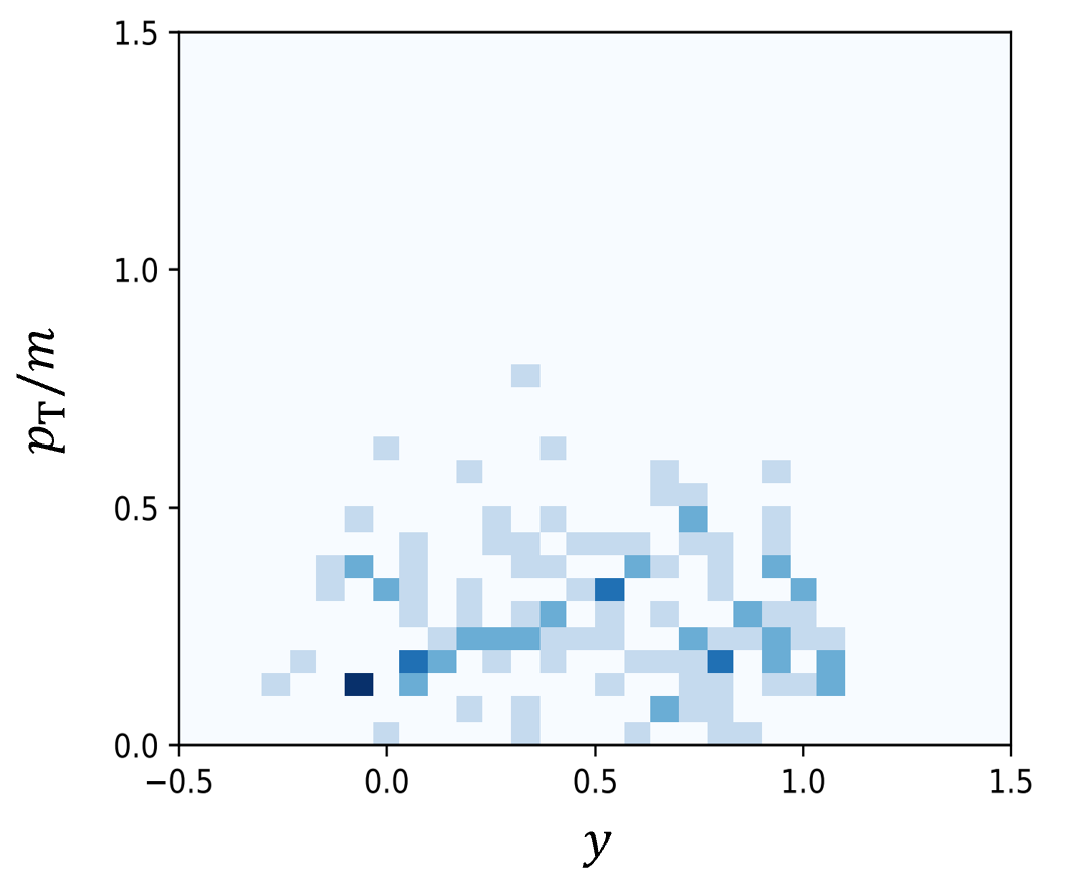}
      \label{fig:6-b}
  }
  \caption{(a) The phase space images in CEE acceptance as a function of the impact parameter. Each image is filled with 10k events with the impact parameter shown in the legend. The binning of the image is 30$\times$30. (b) Image of an event with $b=6$ fm.}
  \label{fig:6}
\end{figure}

We generate the \textbf{phase space dataset with 30$\times$30 binning} to compare the model performance with other work. The range of the phase space used for binning is defined as $y_0\in$[-0.5,1.5], $p_{\rm T}/m\in$[0,1.5]. Figure~\ref{fig:6-a} shows the cumulative images of the phase space as a function of the impact parameter, where the evolution of the image shape can be easily recognized. However, for a single event, features are less evident due to event-by-event fluctuations, as demonstrated in Figure~\ref{fig:6-b}. Two sets of phase space data, with acceptances covered by CEE and only eTOF, have been prepared for a comparison study.

\subsection{Regression models}
Two DNN-based models have been implemented: CNN and Graph Attention network (GAT)\cite{velickovic_graph_2018,brody2022attentive}. The structures of the two models are demonstrated in the Appendix Figure~\ref{fig:a1} and~\ref{fig:a2}. For both models, a dense layer is implemented at the entrance of the input data to encode the three features for each readout strip (762$\times$3). The encoder outputs 762$\times$64 hidden features for further extractions with two sequential CNN or GAT layers. A final dense layer flattens all the output features and decodes them to the impact parameter prediction.

CNN extracts the correlation between geometrically local features and trainable kernels through convolution, and it is one of the most popular algorithms in image recognition. For this work, the phase space distribution can be treated as an image and is capable of inputs with multi-features, similar to the RGB channels of images. Besides the phase space input, CNN can also process digitized signal input. In this mode, features from the 762 strips are arranged into a 96$\times$7$\times$3 shape to reflect the geometry layout. The convolutional kernel in CNN is a crucial module for feature extraction, and its size should be modified according to the geometry coverage of the input data. The kernel size of the first convolutional layer is set to 5$\times$5 for phase space input, while for digi input it is 32$\times$3 because of the strip dimension. The other hyper-parameters of the models can be found in the Appendix Table~\ref{tab:a1}.

Graph Neural Networks (GNN) have recently become a popular machine learning model that can handle graphic structures. For nuclear collisions, data of the produced tracks, hits, or even signals can be regarded as graphs, in which each individual, or node in the concept of a graph, is potentially connected with others under the collision dynamics. Implementing GNNs through the GAT architecture enables robustness in dynamic graphs. It trains a universal and trainable agent for calculating the attention weights between any pair of nodes. Then, the node features can be transferred as a combination of themselves and those of its neighboring nodes, according to the attention weights. Equation~\ref{eq:1a} and~\ref{eq:1b} describes the principle of the GAT layer\footnote{In this work, GATv2 structure\cite{brody2022attentive} is implemented for better reported performance over GAT.}. Firstly, the attention weights $\alpha_{ij}$ are calculated, where $\vec{h}_i$ stands for the input features of node $i$, $\sigma$ the activation function, $\boldmath W$ and $\vec{a}$ the trainable weight matrix and vector. Then, the output features $\vec{h}_i'$ is obtained as the attention-weighted combination of the encoded input features from the self and neighboring nodes $\mathcal{N}_i$.

\begin{equation}
\alpha_{ij} = \frac{\exp\left(\vec{a}^T\sigma(\left[W\vec{h}_i||W\vec{h}_j\right]\right)}{\sum_{k\in \mathcal{N}_i}\exp\left(\vec{a}^T\sigma\left(\left[W\vec{h}_i||W\vec{h}_k\right]\right)\right)}
\label{eq:1a}
\end{equation}

\begin{equation}
\vec{h}_i' =\sum_{j\in \mathcal{N}_i}\alpha_{ij}{\boldmath W}\vec{h}_j 
\label{eq:1b}
\end{equation}

\subsection{Test of the models}
\label{sec:3-3}
The tests described in this paper, along with their key settings are listed in Table~\ref{tab:1}. They are designed for different purposes of performance evaluation and validation. Firstly, to evaluate the model performances, CNN, GAT, and fully connected (FC) neural networks are implemented to regress the hit data of eTOF. Secondly, to compare the input types, CNN models are trained with phase space data and digitized signal data at the forward angle. Finally, for the discussion of further expansion of acceptance with large angle detectors included, we conduct a comparative study with the CNN model to regress the phase space data of the forward angle and total acceptance of CEE. 

In this work, the graph in the GAT network is constructed by connecting every two readout strips, which leaves the network to learn the edge attributes. GAT has a masked attention feature that eliminates attention weights of some predefined edges. Based on the characteristics, we tested the performances of GAT networks when, for each node, the ones outside the kernel of the CNN (32$\times$3) and the ones within the kernel are masked respectively. The tests, labeled as GAT-HIT-FW-f and GAT-HIT-FW-c, can provide evidence on which nodes, farther or closer, are more important to a specific node.

For all the models, mean square error is used as the loss function in training. An early-stop strategy is implemented to keep the best weights of the DNNs when the performance in the valid data is not improved for 10 epochs. The best weights were reached after 10-40 training epochs, depending on the model. Overfitting is not visible in this work since the data amounts are large enough to represent the overall feature distribution. Moreover, the tests have been evaluated on their stability with different train-test data splits.

\begin{table}[htbp]
\centering
\caption{\label{tab:1} The test settings and regression performances in this work.}
\smallskip
\begin{tabular}{lllllll}
\hline
Test label & Acceptance & Type of data input & Network & MAE {[}fm{]} & MSE {[}fm$^2${]} & $R^2$    \\
\hline
CNN-PS-FW  & Forward    & Phase space        & CNN     & 0.583        & 0.585         & 0.927 \\
CNN-PS-TOT & Overall        & Phase space        & CNN     & 0.269        & 0.108         & 0.987 \\
CNN-HIT-FW & Forward    & Hit                & CNN     & 0.506        & 0.435         & 0.946 \\
GAT-HIT-FW & Forward    & Hit                & GAT     & 0.479        & 0.394         & 0.951 \\
GAT-HIT-FW-f & Forward    & Hit                & GAT     & 0.487        & 0.405         & 0.951 \\
GAT-HIT-FW-c & Forward    & Hit                & GAT     & 0.492        & 0.411         & 0.949 \\
FC-HIT-FW  & Forward    & Hit                & FC      & 0.547        & 0.536         & 0.934 \\
\hline
\end{tabular}
\end{table}

\section{Performances and discussion}
\label{sec:4}
The following indicators evaluate the performances of the models: Mean Absolute Error (MAE), Mean Square Error (MSE), and goodness of fit $R^2$. MAE and MSE are calculated with Equation~\ref{eq:2} and~\ref{eq:3} respectively:

\begin{equation}
\sigma_{\rm MAE} = \frac{1}{n}\sum\limits_n\left|b_{\rm pred}-b_{\rm true}\right|
\label{eq:2}
\end{equation}

\begin{equation}
\sigma_{\rm MSE} = \frac{1}{n}\sum\limits_n\left(b_{\rm pred}-b_{\rm true}\right)^2
\label{eq:3}
\end{equation}

The $R^2$ is defined as follows:
\begin{equation}
R^2 = 1 - \frac{\sum_n\left(b_{\rm pred}-b_{\rm true}\right)^2}{\sum_n\left(b_{\rm pred}-\overline{b}\right)^2}
\end{equation}
where $n$ is the size of the data for validation or test and $\overline{b}$ stands for the average value of the impact parameters.

Table~\ref{tab:1} lists the performances of the models under the $b{\rm d}b$ test set. It is reasonable to find that the CNN-PS-TOT model has the best performance for the input features from the total phase space. It can be treated as an estimation of the determination capability in the condition of the CEE spectrometer and energy. The performance of CNN-PS-FW shows the capability with phase space identified by the forward detectors, and it also serves as a reference model for comparison with others that take the forward TOF data as input. It is inspiring to find that the models with eTOF hit input perform better than the reference model, with decreases of the MAE by 6-18\%. Such an improvement in the HIT-FW tests indicates that the digitized signal data preserve implicit patterns that can be extracted by DNN models in diverse forms. For example, the position information can potentially increase the equivalent granularity of the HIT-FW models, and the time may reveal the secondary particle types. Yet, these features are no longer preserved in the phase space inputs.

With phase space data and CNN-based models, CNN-PS-TOT exhibits superior precision compared to CNN-PS-FW. Notably, this precision improvement, attributed to acceptance, is more significant than data format and DNN model. This phenomena indicates that the prediction uncertainty of the model GAT-HIT-FW is primarily aleatoric rather than epistemic. Further quantitative assessment of the uncertainty sources is not covered in this work which only gives error definitions based on ground truth.

\begin{figure}[htbp]
  \centering
  \subfigure[]{
      \includegraphics[width=0.45\textwidth]{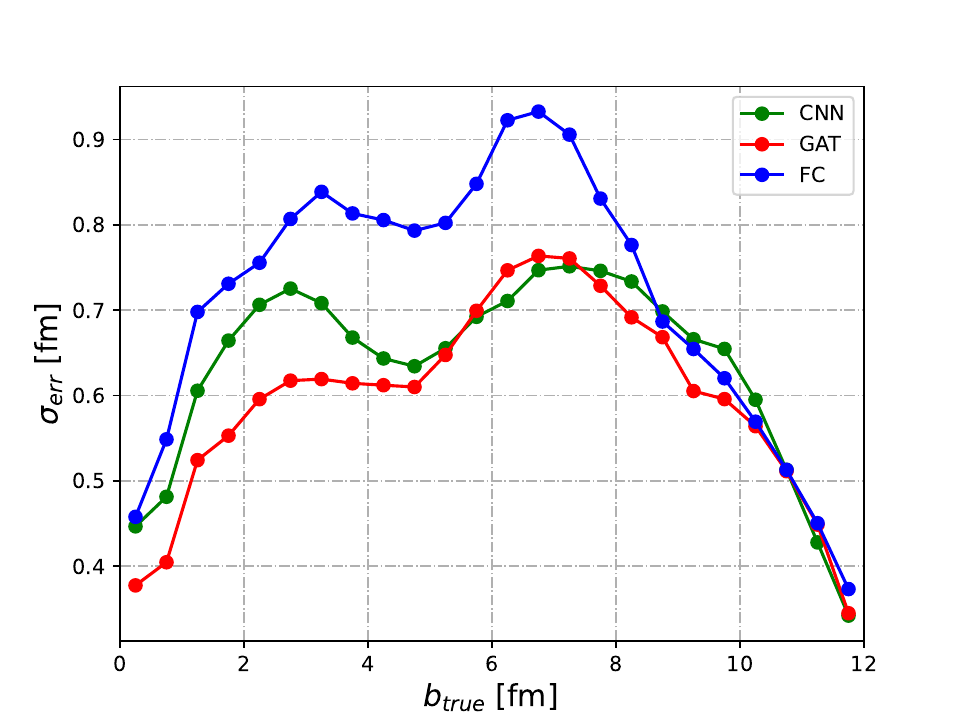}
      \label{fig:7-a}
  }
  % \hfill
  \subfigure[]{
      \includegraphics[width=0.45\textwidth]{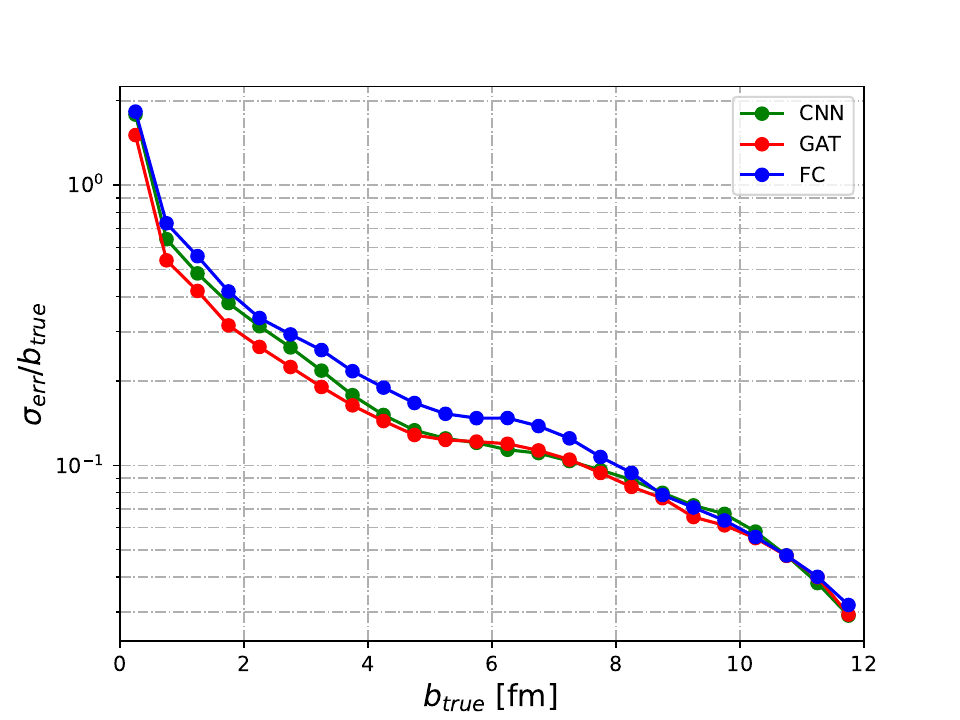}
      \label{fig:7-b}
  }
  \caption{Prediction MAE as a function of impact parameter.}
  \label{fig:7}
\end{figure}

Among the HIT-FW tests, the GAT model has the highest precision in centrality prediction. The dependency of the prediction resolution (MAE) on the impact parameter for the HIT-FW tests is investigated, as shown in Figure~\ref{fig:7}. Each data point is defined on the uniform test set with an impact parameter interval of [-0.25, 0.25) fm. For example, the first point, at 0.25 fm, corresponds to the data in [0, 0.5) fm. The three models show similar patterns in their curves. Still, for most centrality, including the central region, the GAT model outperforms other models significantly, given the statistical uncertainty of 0.4\% (50k events for each data point).

The aggregation mechanism of neighbors in the GAT structure contributes to the overall superiority because it helps eliminate the randomness of the single hits. Such a principle is also embodied in the CNN model, which is based on spatial convolution. However, the convolution kernel size confines the aggregation range. It can be seen that masking either the local (GAT-HIT-FW-f) or the distant nodes (GAT-HIT-FW-c) from transferring their features will decrease the prediction power of the GAT model.

\begin{figure}[htbp]
  \centering
  \subfigure[]{
      \includegraphics[width=0.45\textwidth]{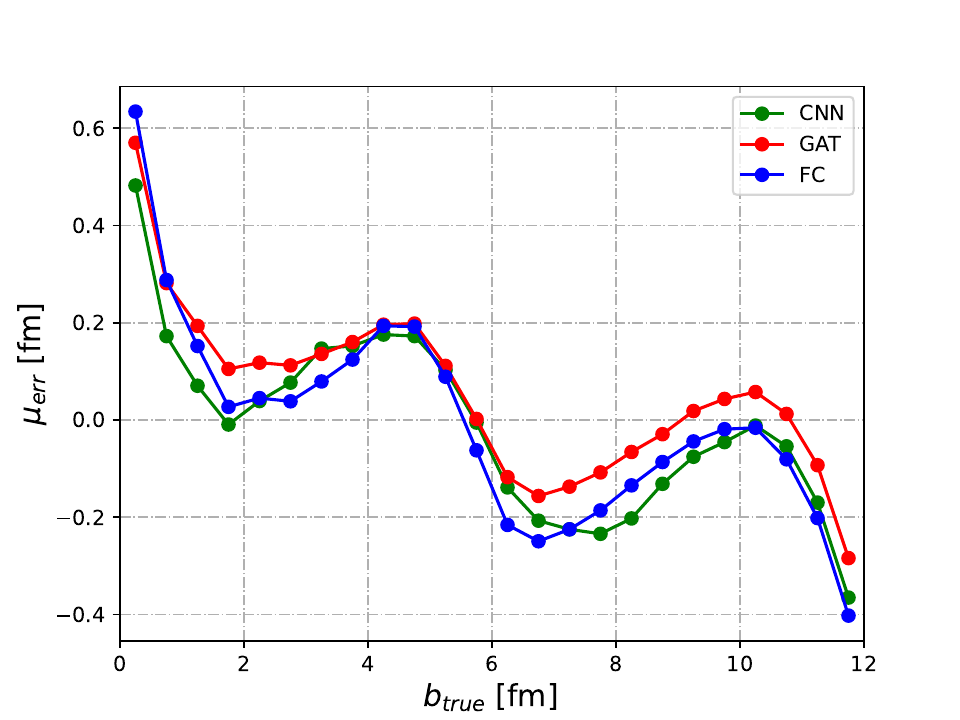}
      \label{fig:8-a}
  }
  % \hfill
  \subfigure[]{
      \includegraphics[width=0.45\textwidth]{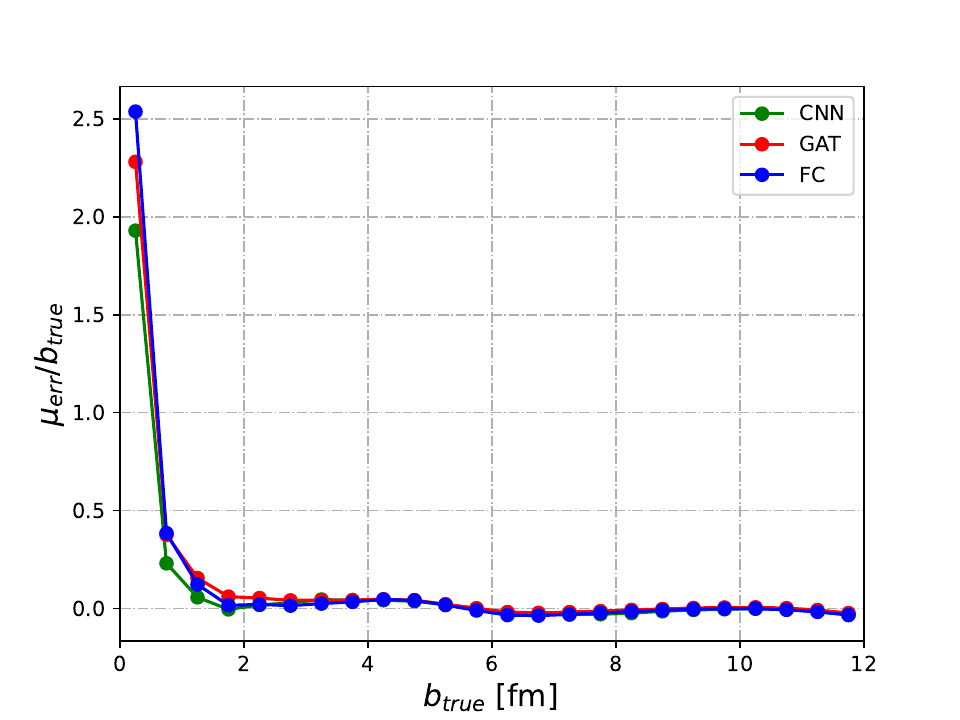}
      \label{fig:8-b}
  }
  \caption{Prediction Bias as a function of the impact parameter.}
  \label{fig:8}
\end{figure}

Figure~\ref{fig:8} shows the relative prediction biases of the HIT-FW tests as a function of centrality. The three models show satisfying bias suppression. For very central collisions (b<1.5 fm), the relative bias rises significantly. The lowest accuracy of the GAT model will not significantly influence its application, given the best prediction MAE which has taken the bias into account.

\begin{figure}[htbp]
\centering
\includegraphics[width=.7\textwidth]{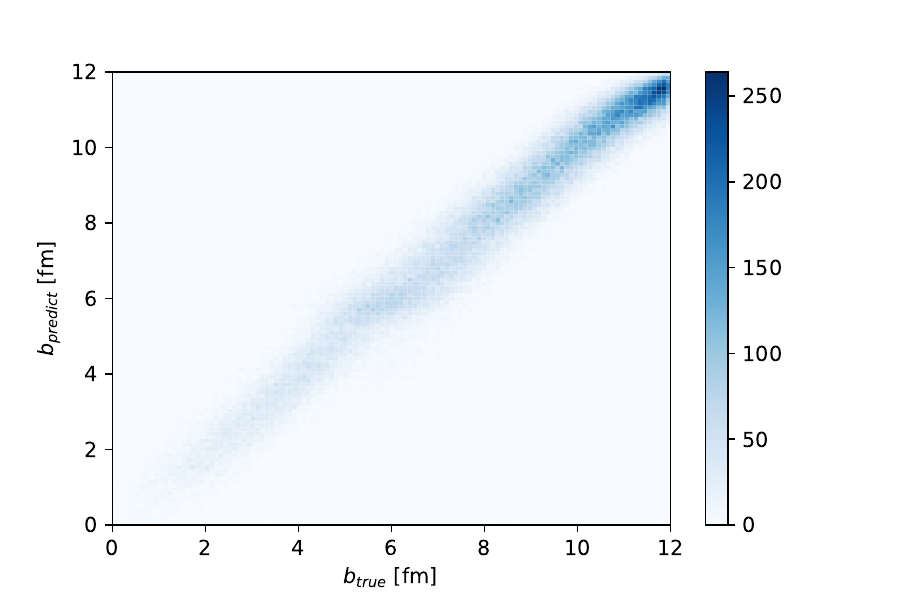}
\caption{\label{fig:9} Distribution between the predicted impact parameters by GAT-HIT-FW and the true impact parameters.}
\end{figure}

\begin{figure}[htbp]
\centering
\includegraphics[width=.95\textwidth]{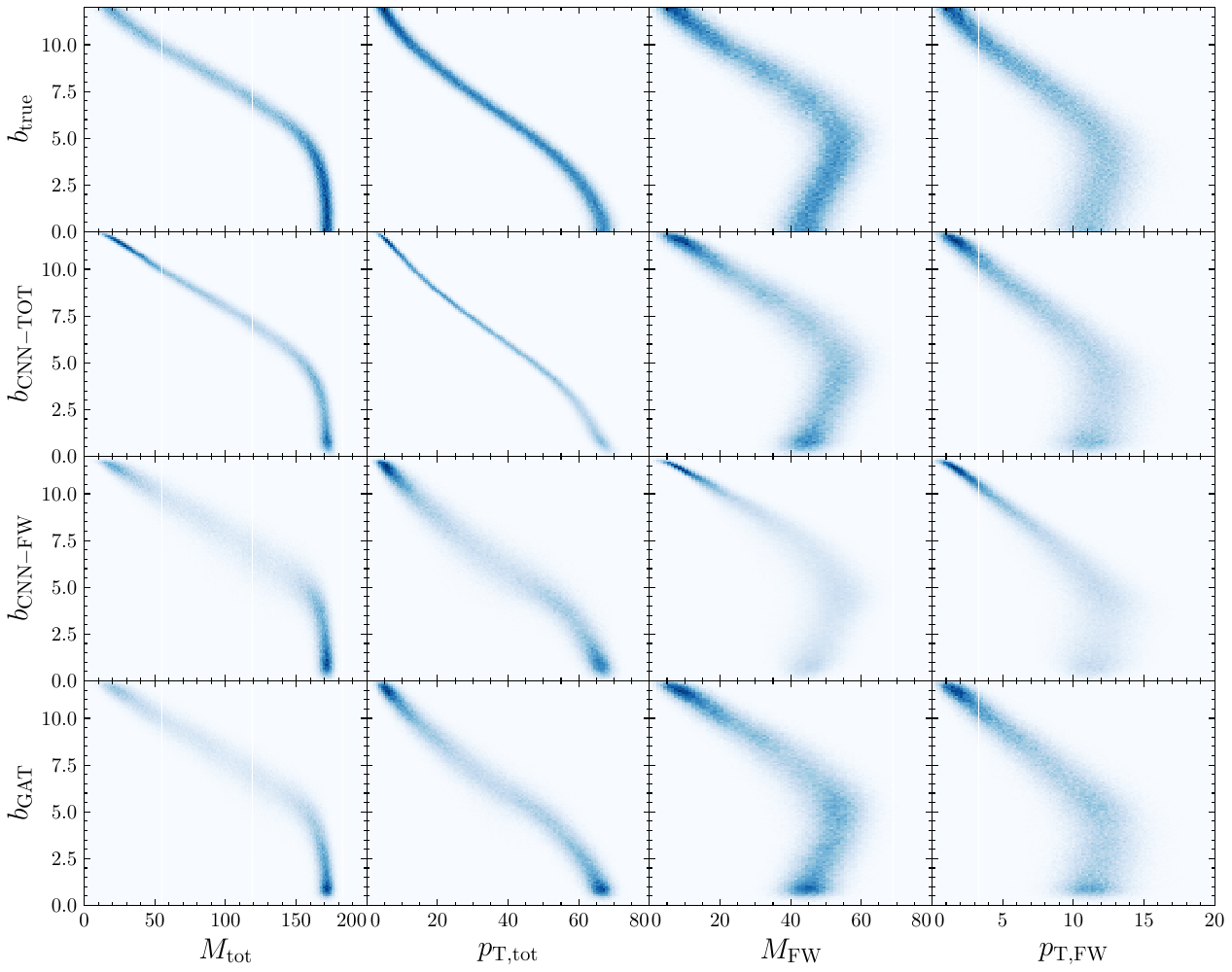}
\caption{\label{fig:10} Dependency of the true and predicted impact parameters on the correlated collision features.}
\end{figure}

The prediction of the GAT model on the $b{\rm d} b$ test dataset is examined visually in Figure~\ref{fig:9}. A good correlation is shown, which agrees with the results in Table~\ref{tab:1}. The linear relationship seems to be lost at around 6 fm, where the prediction error is higher than in other ranges in Figure~\ref{fig:7}. Similar patterns are visible for other models as well. Figure~\ref{fig:10} shows the dependency of impact parameters on the correlated collision features. $b_{\rm CNN-TOT}$, $b_{\rm CNN-FW}$, $b_{\rm GAT}$ represent the predicted impact parameters from model CNN-PS-TOT, CNN-PS-FW and GAT-HIT-FW respectively, $M_{\rm tot}$ the multiplicity of total acceptance, $p_{\rm T, tot}$ the summation of reduced transverse momentum for charged particles in the overall acceptance, $M_{\rm fw}$, $p_{\rm T, fw}$ the ones in the forward acceptance. It is shown that the correlation of the features at forward acceptance is weak for impact parameters around 6 fm. This may explain the low precision of such a region for forward predictors. When the impact parameter decreases, the secondary particle distribution moves toward the mid-rapidity region, which goes beyond the coverage of eTOF. Consequently, the forward quantities lose the monotonicity and decline, which supports why the model fed with overall phase spaces performs best. In general, the models trained from either the phase spaces or the digitized signals successfully reproduce the correlations between the impact parameter and the collision quantities.

\begin{figure}[htbp]
\centering
\includegraphics[width=.5\textwidth]{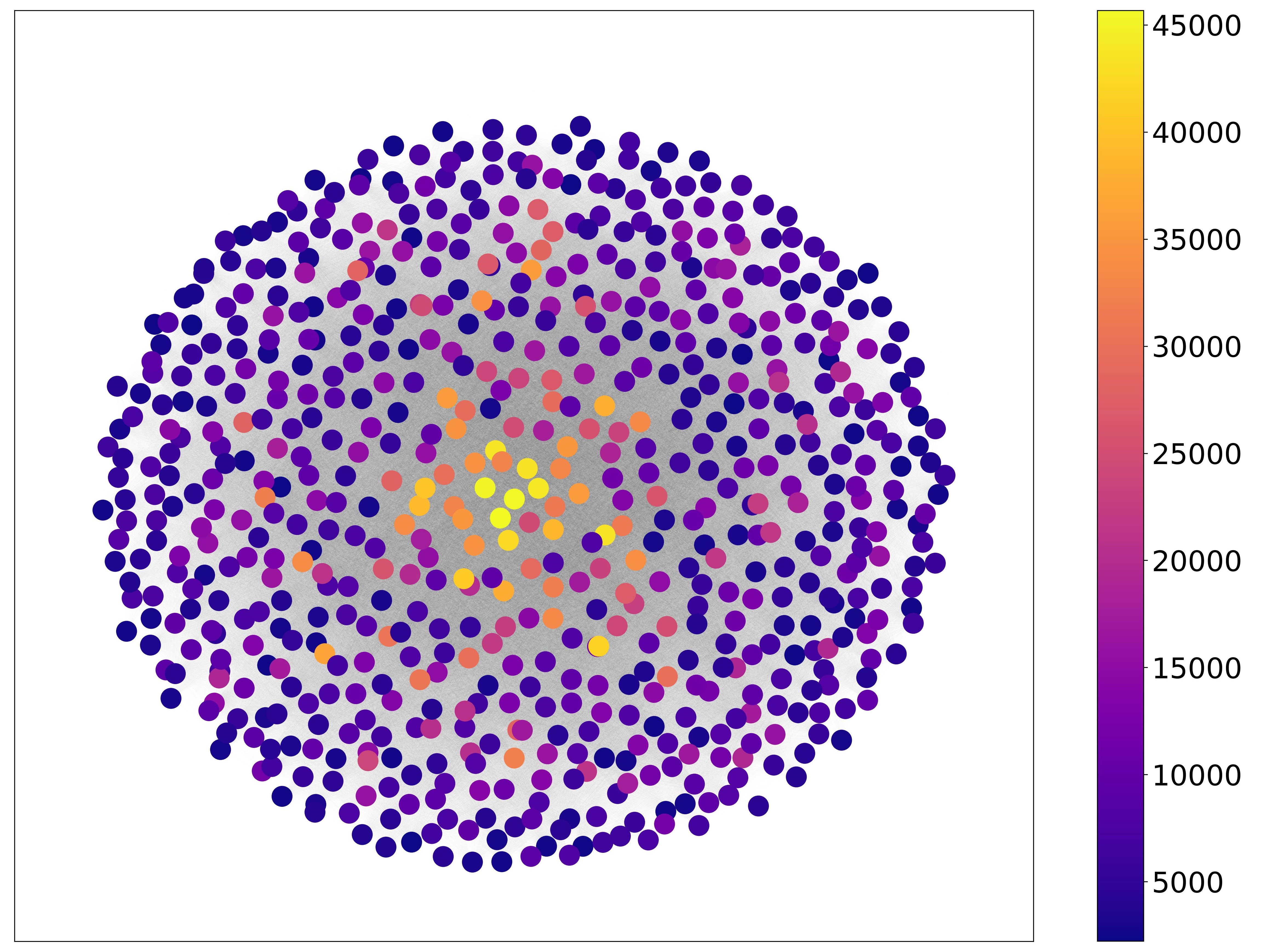}
\caption{\label{fig:11} Visualization of the edge weights between 672 readout strips after training.}
\end{figure}

A significant characteristic of the GAT model is the interpretability of graphical relations. We investigate the importance of the attention weights among the 672 readout strips with the help of NetworkX\cite{noauthor_networkx_nodate}, a tool for visualizing and analyzing the graph. Figure~\ref{fig:11} shows the graph visualization of the GAT model in predicting the test dataset. The grey cloud includes an intense crowd of black edges whose widths represent the average value of the attention weights. The nodes in the graph are arranged following the Fruchterman-Reingold force-directed algorithm\cite{fruchterman_graph_1991}, and as a result, nodes with stronger connections to others will be dragged to the central area. The nodes in the graph are colored according to the total signal counts on the strip. It is observed that the light-colored nodes, which represent the strips from the inner area of eTOF, have a higher importance in the contribution of the features, which indicates that the features from these strips can be effectively transmitted to the other strips, even to those at the outer region.

\section{Conclusion}
This work is one of the efforts to determine the impact parameter for CEE. DNN-based models are regarded as appropriate approaches as high dimensional features are needed in low collision energies. We propose that the original digitized signals from the eTOF MRPCs are informative as input and that the DNN models are powerful enough to extract the predictive features. Several structures of DNN models are implemented and tested, showing that the prediction precision with digitized signal inputs exceeds the phase space inputs of the same acceptance by up to 18\%. Among the models in our tests, the GAT model reaches the best prediction precision of 0.479 fm in mean absolute error, and the prediction power outperforms other models significantly for central collisions. It is also examined that larger acceptances will result in better performance, which encourages further investigations of including data from other subsystems.

\acknowledgments

This work is supported by the National Natural Science Foundation of China under Grant No. 11927901, 11420101004, 11735009, U1832118, and by the Ministry of Science and Technology under Grant No. 2018YFE0205200, 2016YFA0400100.

\appendix
\section{Structures and hyper-parameters of the models}

\begin{figure}[htbp]
\centering
\includegraphics[width=.8\textwidth]{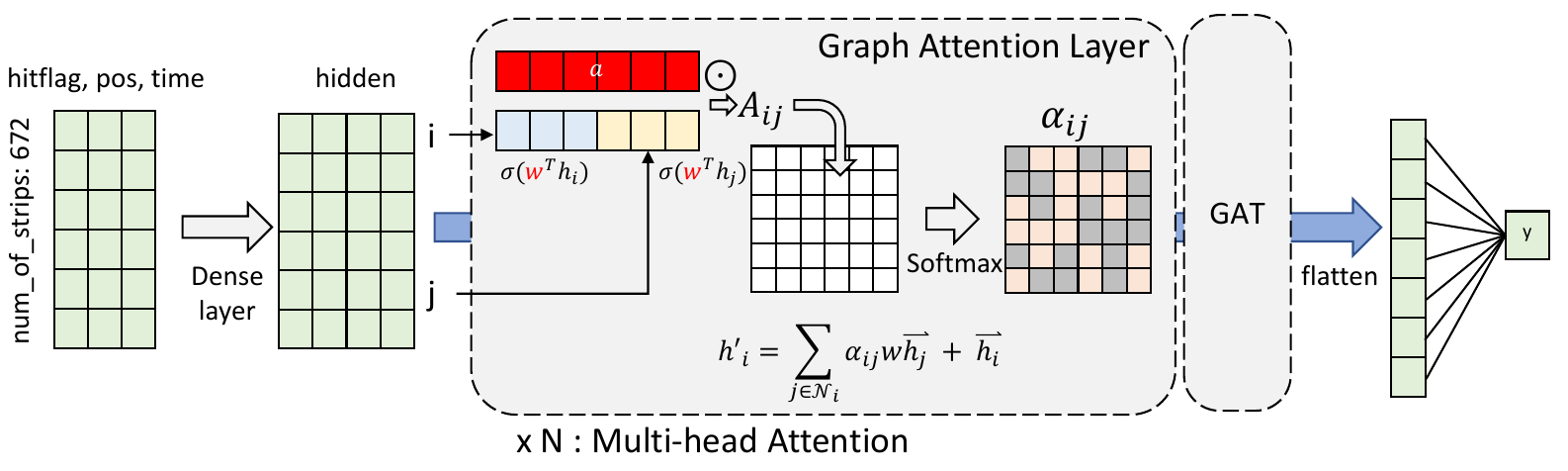}
\caption{\label{fig:a1} Structure of the GAT network used in this work.}
\end{figure}

\begin{figure}[htbp]
\centering
\includegraphics[width=.8\textwidth]{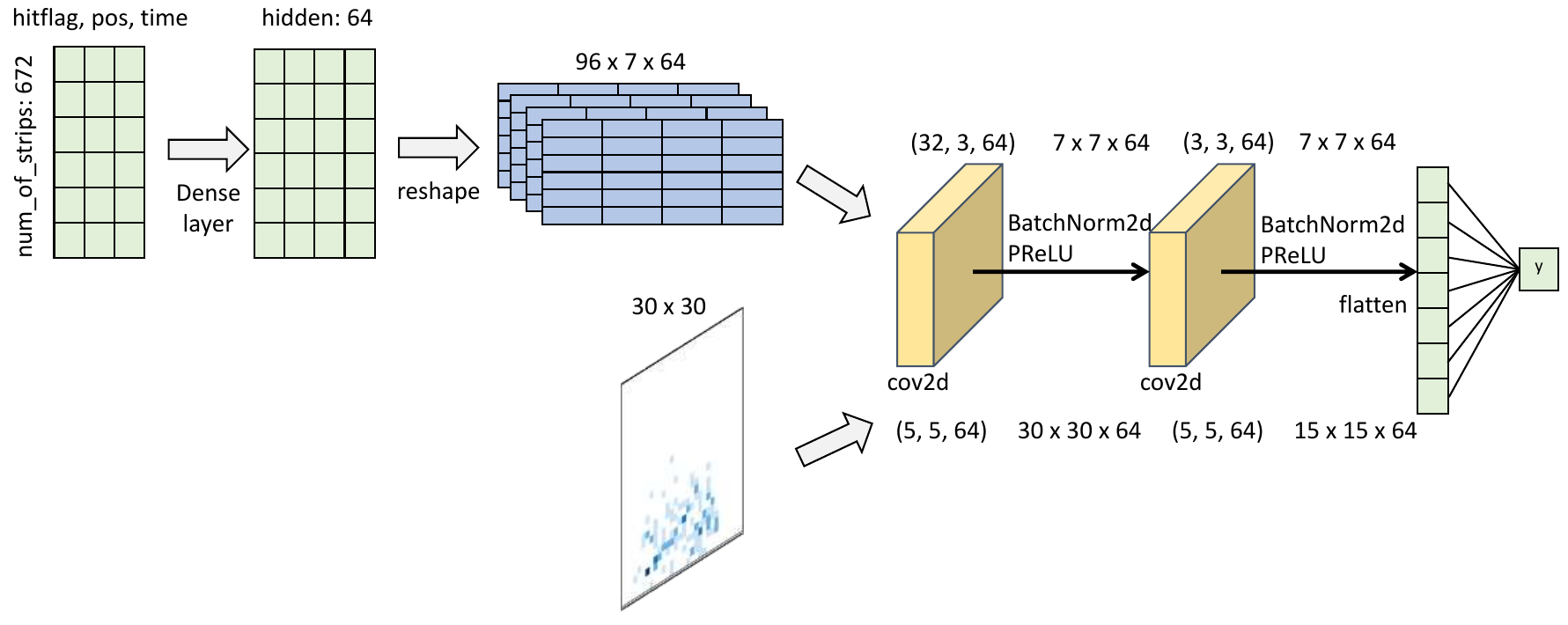}
\caption{\label{fig:a2} Structure of the CNN network used in this work. The upper flow processes digitized signal inputs, while the lower flow processes phase space inputs.}
\end{figure}

\begin{table}[htbp]
\centering
\caption{Hyperparameters for the models in this work.}
\begin{tabular}{lllllll}
\hline
\multicolumn{2}{c}{GAT}                                 & \multicolumn{2}{c}{FC}                     &  &  &  \\
\hline
Hidden size            & 64 into 4 heads                 & Hidden size            & 64                &  &  &  \\
Num. layers             & 2                              & Num. layers            & 2                 &  &  &  \\
Learning rate          & 5$\times$10$^{-4}$                         & Learning rate          & 5$\times$10$^{-4}$            &  &  &  \\
Learning rate strategy & Warmup and exponential   decay & Learning rate strategy & Exponential decay &  &  &  \\
\hline
\multicolumn{2}{c}{CNN-HIT}                             & \multicolumn{2}{c}{CNN-PS}                 &  &  &  \\
\hline
Hidden channels        & 64                             & Hidden channels        & 64                &  &  &  \\
Num. conv. layers      & 2                              & Num. conv. layers      & 2                 &  &  &  \\
Kernel size 1          & (3, 32, 64)                    & Kernel size 1          & (5, 5, 64)        &  &  &  \\
Kernel size 2          & (3, 3, 64)                     & Kernel size 2          & (5, 5, 64)        &  &  &  \\
Learning rate          & 5$\times$10$^{-4}$                         & Learning rate          & 5$\times$10$^{-4}$            &  &  &  \\
Learning rate strategy & Exponential decay              & Learning rate strategy & Exponential decay &  &  & \\
\hline
\end{tabular}
\label{tab:a1}
\end{table}

% Please always give a title also for appendices.

% \paragraph{Note added.} This is also a good position for notes added
% after the paper has been written.


% \begin{thebibliography}{99}

% \bibitem{a}
% Author, \emph{Title}, \emph{J. Abbrev.} {\bf vol} (year) pg.

% \bibitem{b}
% Author, \emph{Title},
% arxiv:1234.5678.

% \bibitem{c}
% Author, \emph{Title},
% Publisher (year).

% % Please avoid comments such as "For a review'', "For some examples",
% % "and references therein" or move them in the text. In general,
% % please leave only references in the bibliography and move all
% % accessory text in footnotes.

% % Also, please have only one work for each \bibitem.

% \end{thebibliography}

\bibliographystyle{JHEP}
\bibliography{refs}

\providecommand{\href}[2]{#2}\begingroup\raggedright\begin{thebibliography}{10}

\bibitem{jacob_quark_1982}
M.~Jacob and J.~Tran Thanh~Van, \emph{Quark matter formation and heavy ion
  collisions: {A} general review and status report},
  \href{https://doi.org/10.1016/0370-1573(82)90083-7}{\emph{Phys. Rep.}
  {\bfseries 88} (1982) 321}.

\bibitem{gupta_scale_2011}
S.~Gupta et~al., \emph{Scale for the {Phase} {Diagram} of {Quantum}
  {Chromodynamics}}, \href{https://doi.org/10.1126/science.1204621}{\emph{Sci}
  {\bfseries 332} (2011) 1525}.

\bibitem{luo_search_2017}
X.~Luo and N.~Xu, \emph{Search for the {QCD} critical point with fluctuations
  of conserved quantities in relativistic heavy-ion collisions at {RHIC}: an
  overview}, \href{https://doi.org/10.1007/s41365-017-0257-0}{\emph{Nucl. Sci.
  Tech.} {\bfseries 28} (2017) 112}.

\bibitem{jacak_exploration_2012}
B.~V. Jacak and B.~Müller, \emph{The {Exploration} of {Hot} {Nuclear}
  {Matter}}, \href{https://doi.org/10.1126/science.1215901}{\emph{Sci}
  {\bfseries 337} (2012) 310}.

\bibitem{alford_color_2008}
M.~G. Alford et~al., \emph{Color superconductivity in dense quark matter},
  \href{https://doi.org/10.1103/RevModPhys.80.1455}{\emph{Rev. Mod. Phys.}
  {\bfseries 80} (2008) 1455}.

\bibitem{ackermann_star_2003}
K.~Ackermann et~al., \emph{{STAR} detector overview},
  \href{https://doi.org/10.1016/S0168-9002(02)01960-5}{\emph{Nucl. Instrum.
  Methods Phys. Res. A} {\bfseries 499} (2003) 624}.

\bibitem{noauthor_alice_2008}
T.~A. Collaboration et~al., \emph{The {ALICE} experiment at the {CERN} {LHC}},
  \href{https://doi.org/10.1088/1748-0221/3/08/S08002}{\emph{J. Inst.}
  {\bfseries 3} (2008) S08002}.

\bibitem{kim_detector_2019}
Y.~Kim, \emph{The {Detector} {Development} and {Physics} {Program} in {sPHENIX}
  {Experiment} at {RHIC}},
  \href{https://doi.org/10.1016/j.nuclphysa.2018.10.075}{\emph{Nucl. Phys. A}
  {\bfseries 982} (2019) 955}.

\bibitem{meehan_fixed_2016}
K.~C. Meehan, \emph{Fixed {Target} {Collisions} at {STAR}},
  \href{https://doi.org/10.1016/j.nuclphysa.2016.04.016}{\emph{Nucl. Phys. A}
  {\bfseries 956} (2016) 878}.

\bibitem{galatyuk_hades_2014}
T.~Galatyuk, \emph{{HADES} overview},
  \href{https://doi.org/10.1016/j.nuclphysa.2014.10.044}{\emph{Nucl. Phys. A}
  {\bfseries 931} (2014) 41}.

\bibitem{lu_conceptual_2017}
L.~Lü et~al., \emph{Conceptual design of the {HIRFL}-{CSR} external-target
  experiment}, \href{https://doi.org/10.1007/s11433-016-0342-x}{\emph{Sci.
  China Phys., Mech. Astron.} {\bfseries 60} (2017) 012021}.

\bibitem{wang_multigap_2020}
Y.~Wang and Y.~Yu, \emph{Multigap {Resistive} {Plate} {Chambers} for {Time} of
  {Flight} {Applications}},
  \href{https://doi.org/10.3390/app11010111}{\emph{Appl. Sci.} {\bfseries 11}
  (2020) 111}.

\bibitem{huang_laser_2018}
W.~Huang et~al., \emph{Laser test of the prototype of {CEE} time projection
  chamber}, \href{https://doi.org/10.1007/s41365-018-0382-4}{\emph{Nucl. Sci.
  Tech.} {\bfseries 29} (2018) 41}.

\bibitem{wang_cee_2022}
X.~Wang et~al., \emph{{CEE} inner {TOF} prototype design and preliminary test
  results}, \href{https://doi.org/10.1088/1748-0221/17/09/P09023}{\emph{J.
  Instrum.} {\bfseries 17} (2022) P09023}.

\bibitem{yi_prototype_2014}
H.~Yi et~al., \emph{Prototype studies on the forward {MWDC} tracking array of
  the external target experiment at {HIRFL}-{CSR}},
  \href{https://doi.org/10.1088/1674-1137/38/12/126002}{\emph{Chin. Phys. C}
  {\bfseries 38} (2014) 126002}.

\bibitem{wang_cee-etof_2020}
B.~Wang et~al., \emph{The {CEE}-{eTOF} wall constructed with new sealed
  {MRPC}}, \href{https://doi.org/10.1088/1748-0221/15/08/C08022}{\emph{J.
  Instrum.} {\bfseries 15} (2020) C08022}.

\bibitem{wang_external_2023}
B.~Wang et~al., \emph{The external time-of-flight wall for {CEE} experiment},
  \href{https://doi.org/10.1140/epjc/s10052-023-11806-2}{\emph{Eur. Phys. J. C}
  {\bfseries 83} (2023) 817}.

\bibitem{hu_extensive_2020}
D.~Hu et~al., \emph{Extensive beam test study of prototype {MRPCs} for the {T0}
  detector at the {CSR} external-target experiment},
  \href{https://doi.org/10.1140/epjc/s10052-020-7804-2}{\emph{Eur. Phys. J. C}
  {\bfseries 80} (2020) 282}.

\bibitem{liu_event_2023}
L.-K. Liu, H.~Pei, Y.-P. Wang, B.~Zhang, N.~Xu and S.-S. Shi, \emph{Event plane
  determination from the zero degree calorimeter at the cooling storage ring
  external-target experiment},
  \href{https://doi.org/10.1007/s41365-023-01262-8}{\emph{NUCL SCI TECH}
  {\bfseries 34} (2023) 100}.

\bibitem{goyal_impact_2013}
S.~Goyal, \emph{Impact parameter dependence of collective flow and its
  disappearance for different mass asymmetries},
  \href{https://doi.org/10.1140/epja/i2013-13153-1}{\emph{Eur. Phys. J. A}
  {\bfseries 49} (2013) 153}.

\bibitem{kaur_elliptical_2011}
V.~Kaur et~al., \emph{On the elliptical flow and mass asymmetry of the
  colliding nuclei},
  \href{https://doi.org/10.1016/j.physletb.2011.02.044}{\emph{Phys. Lett. B}
  {\bfseries 697} (2011) 512}.

\bibitem{cavata_determination_1990}
C.~Cavata et~al., \emph{Determination of the impact parameter in relativistic
  nucleus-nucleus collisions},
  \href{https://doi.org/10.1103/PhysRevC.42.1760}{\emph{Phys. Rev. C}
  {\bfseries 42} (1990) 1760}.

\bibitem{miller_glauber_2007}
M.~L. Miller et~al., \emph{Glauber {Modeling} in {High}-{Energy} {Nuclear}
  {Collisions}},
  \href{https://doi.org/10.1146/annurev.nucl.57.090506.123020}{\emph{Annu. Rev.
  Nucl. Part. S.} {\bfseries 57} (2007) 205}.

\bibitem{frankland_model_2021}
J.~D. Frankland et~al., \emph{Model independent reconstruction of impact
  parameter distributions for intermediate energy heavy ion collisions},
  \href{https://doi.org/10.1103/PhysRevC.104.034609}{\emph{Phys. Rev. C}
  {\bfseries 104} (2021) 034609}.

\bibitem{das_relating_2018}
S.~J. Das et~al., \emph{Relating centrality to impact parameter in
  nucleus-nucleus collisions},
  \href{https://doi.org/10.1103/PhysRevC.97.014905}{\emph{Phys. Rev. C}
  {\bfseries 97} (2018) 014905}.

\bibitem{de_sanctis_classification_2009}
J.~De~Sanctis et~al., \emph{Classification of the impact parameter in
  nucleus–nucleus collisions by a support vector machine method},
  \href{https://doi.org/10.1088/0954-3899/36/1/015101}{\emph{J. Phys. G: Nucl.
  Part. Phys.} {\bfseries 36} (2009) 015101}.

\bibitem{li_application_2020}
F.~Li et~al., \emph{Application of artificial intelligence in the determination
  of impact parameter in heavy-ion collisions at intermediate energies},
  \href{https://doi.org/10.1088/1361-6471/abb1f9}{\emph{J. Phys. G: Nucl. Part.
  Phys.} {\bfseries 47} (2020) 115104}.

\bibitem{haddad_impact_1997}
F.~Haddad et~al., \emph{Impact parameter determination in experimental analysis
  using a neural network},
  \href{https://doi.org/10.1103/PhysRevC.55.1371}{\emph{Phys. Rev. C}
  {\bfseries 55} (1997) 1371}.

\bibitem{david_impact_1995}
C.~David, M.~Freslier and J.~Aichelin, \emph{Impact parameter determination for
  heavy-ion collisions by use of a neural network},
  \href{https://doi.org/10.1103/PhysRevC.51.1453}{\emph{Phys. Rev. C}
  {\bfseries 51} (1995) 1453}.

\bibitem{omana_kuttan_fast_2020}
M.~Omana~Kuttan et~al., \emph{A fast centrality-meter for heavy-ion collisions
  at the {CBM} experiment},
  \href{https://doi.org/10.1016/j.physletb.2020.135872}{\emph{Phys. Lett. B}
  {\bfseries 811} (2020) 135872}.

\bibitem{krizhevsky_imagenet_2017}
A.~Krizhevsky, I.~Sutskever and G.~E. Hinton, \emph{{ImageNet} classification
  with deep convolutional neural networks},
  \href{https://doi.org/10.1145/3065386}{\emph{Commun. ACM} {\bfseries 60}
  (2017) 84}.

\bibitem{li_application_2021}
F.~Li et~al., \emph{Application of machine learning in the determination of
  impact parameter in the {Sn} 132 + {Sn} 124 system},
  \href{https://doi.org/10.1103/PhysRevC.104.034608}{\emph{Phys. Rev. C}
  {\bfseries 104} (2021) 034608}.

\bibitem{anghinolfi_nino_2004}
F.~Anghinolfi et~al., \emph{{NINO}: an ultra-fast and low-power front-end
  amplifier/discriminator {ASIC} designed for the multigap resistive plate
  chamber}, \href{https://doi.org/10.1016/j.nima.2004.07.024}{\emph{Nucl.
  Instrum. Meth A} {\bfseries 533} (2004) 183}.

\bibitem{lu_readout_2021}
J.~Lu et~al., \emph{Readout {Electronics} {Prototype} of {TOF} {Detectors} in
  {CEE} of {HIRFL}}, \href{https://doi.org/10.1109/TNS.2021.3093544}{\emph{IEEE
  Trans. Nucl. Sci.} {\bfseries 68} (2021) 1976}.

\bibitem{hartnack_modelling_1998}
C.~Hartnack et~al., \emph{Modelling the many-body dynamics of heavy ion
  collisions: {Present} status and future perspective},
  \href{https://doi.org/10.1007/s100500050045}{\emph{Eur. Phys. J. - Hadrons
  Nucl.} {\bfseries 1} (1998) 151}.

\bibitem{noauthor_ceeroot_nodate}
\emph{{CeeRoot}: {CEERoot} simulation and analysis package},
  \href{https://gitee.com/CEESM/CeeRoot}{https://gitee.com/CEESM/CeeRoot}.

\bibitem{al-turany_fairroot_2012}
M.~Al-Turany et~al., \emph{The {FairRoot} framework},
  \href{https://doi.org/10.1088/1742-6596/396/2/022001}{\emph{J. Phys.: Conf.
  Ser.} {\bfseries 396} (2012) 022001}.

\bibitem{wang_simulation_2022}
B.~Wang et~al., \emph{A simulation and analysis framework for {CEE}-{eTOF}},
  \href{https://doi.org/10.1088/1748-0221/17/07/P07024}{\emph{J. Inst.}
  {\bfseries 17} (2022) P07024}.

\bibitem{velickovic_graph_2018}
P.~Veličković et~al., \emph{Graph attention networks},  2018,
  \href{https://arxiv.org/abs/1710.10903}{{\ttfamily 1710.10903}}.

\bibitem{brody2022attentive}
S.~Brody, U.~Alon and E.~Yahav, \emph{How attentive are graph attention
  networks?},  2022, \href{https://arxiv.org/abs/2105.14491}{{\ttfamily
  2105.14491}}.

\bibitem{noauthor_networkx_nodate}
\emph{{NetworkX} — {NetworkX} documentation},
  \href{https://networkx.org/}{https://networkx.org/}.

\bibitem{fruchterman_graph_1991}
T.~M.~J. Fruchterman and E.~M. Reingold, \emph{Graph drawing by force-directed
  placement}, \href{https://doi.org/10.1002/spe.4380211102}{\emph{Softw.:
  Pract. Exper.} {\bfseries 21} (1991) 1129}.

\end{thebibliography}\endgroup

\end{document}